\documentclass[a4paper,fleqn,usenatbib]{mnras}

\usepackage[T1]{fontenc}
\usepackage{ae,aecompl}

\usepackage{graphicx}	
\usepackage{amsmath}	
\usepackage{amssymb}	

\usepackage[normalem]{ulem}

\title[Effective description of DM self-interactions]{Effective description of dark matter self-interactions in small dark matter haloes\thanks{Preprint number: DESY-17-089}}

\author[J.~Kummer, F.~Kahlhoefer and K.~Schmidt-Hoberg]
  {Janis Kummer$^{1,2}$\textsuperscript{\thanks{janis.kummer@desy.de}},
    Felix Kahlhoefer$^{1,3}$ and Kai Schmidt-Hoberg$^1$
\\
$^1$ DESY, Notkestrasse 85, D-22607 Hamburg, Germany\\
$^2$ Hamburger Sternwarte, Gojenbergsweg 112, D-21029 Hamburg, Germany\\
$^3$ Institute for Theoretical Particle Physics and Cosmology (TTK), RWTH Aachen University, D-52056 Aachen, Germany\\}

\date{Accepted XXX. Received YYY; in original form ZZZ}

\pubyear{2017}

\begin{document}
\label{firstpage}
\pagerange{\pageref{firstpage}--\pageref{lastpage}}
\maketitle

\begin{abstract}
Self-interacting dark matter may have striking astrophysical signatures, such as observable offsets between galaxies and dark matter in merging galaxy clusters.
Numerical $N$-body simulations used to predict such observables typically treat the galaxies as collisionless test particles, a questionable assumption given that each galaxy is embedded in its own dark matter halo. To enable a more accurate treatment we develop an effective description of small dark matter haloes taking into account the two major effects due to dark matter self-scatterings: deceleration and evaporation. We point out that self-scatterings can have a sizeable impact on the trajectories of galaxies, diminishing the separation between galaxies and dark matter in merging clusters. This effect depends sensitively on the underlying particle physics, in particular the angular dependence of the self-scattering cross section, and cannot be predicted from the momentum transfer cross section alone.
\end{abstract}

\begin{keywords}
astroparticle physics -- dark matter -- galaxies: clusters: general
\end{keywords}



\section{Introduction}

Although gravitational probes at all astrophysical scales yield strong evidence for the existence of dark matter (DM), its particle physics nature and in particular its interactions remain unknown. Laboratory experiments constrain the couplings of DM to ordinary matter to be very small, but they cannot exclude the possibility that DM has large self-interactions. In fact, self-interacting dark matter (SIDM) is not only well-motivated from the particle physics perspective~\citep{Carlson:1992fn,Kusenko:2001vu,Mohapatra:2001sx}, but it may even provide a solution to the well-known small scale problems of the $\Lambda$CDM paradigm \citep{deLaix:1995vi,Spergel:1999mh}.\footnote{For further details, we refer to a detailed recent review by~\cite{Tulin:2017ara}.}

It is however difficult to infer the role of DM self-interactions from these observations alone, given alternative solutions to the small-scale problems such as baryonic feedback processes~\citep{Navarro:1996bv,
Governato:2009bg}. Establishing the effects of DM self-interactions on astrophysical scales will therefore require a number of different observables. A particularly conclusive signature of SIDM would be the observation of an offset between the stars or galaxies and the centroid of the corresponding DM halo of a system moving through a background DM density~\citep{Williams:2011pm,Dawson:2011kf,Harvey:2013bfa,Harvey:2013tfa,Kahlhoefer:2013dca,Mohammed:2014iya,Harvey:2015hha,Harvey:2016bqd,Kim:2016ujt,Robertson:2016qef,Wittman:2017gxn}. Such a system could for example be a galaxy falling towards the centre of a galaxy cluster, as recently observed within A3827 \citep{Massey:2015dkw,Kahlhoefer:2015vua}, or a merging galaxy cluster like the `Bullet Cluster'~\citep{Markevitch:2003at,Clowe:2006eq,Randall:2007ph}. The former case is conceptually simple, because the luminous matter component comprises individual stars, which are clearly unaffected by DM self-interactions. In this article, however, we focus on the latter case, where one needs to compare the DM distribution of the galaxy cluster with the distribution of galaxies, which have their own DM haloes and which therefore may also be affected by SIDM.

To make predictions for the effects of DM self-interactions in such environments and to interpret observations, highly refined $N$-body simulations of SIDM have been developed~\citep{Randall:2007ph,Vogelsberger:2012ku,Zavala:2012us,Peter:2012jh,Rocha:2012jg,Vogelsberger:2014pda,Cyr-Racine:2015ihg,Vogelsberger:2015gpr,Kim:2016ujt,Robertson:2016qef}. In state-of-the-art simulations it is possible in principle to simulate a merging galaxy cluster with sufficient resolution to resolve the individual galaxies and consistently include DM self-interactions. However, the relatively small number of galaxies in a galaxy cluster leads to large fluctuations in the predicted offset between the galaxies and the DM halo as well as a strong sensitivity to initial conditions.

A possible way to resolve this issue and extract statistically accurate predictions would be to perform a large number of simulations with different initial conditions. A computationally much less expensive and therefore more popular approach to solve this problem is to simulate a large number of effective galaxy particles~\citep{Randall:2007ph, Kim:2016ujt,Robertson:2016xjh}, which have much smaller masses than typical galaxies but are much more abundant. Moreover past and current $N$-body simulations assume that these test galaxies can be treated as collisionless test particles even in the presence of DM self-interactions. It should be clear however that these test galaxies are no longer a faithful representation of real galaxies and major effects may be missed due to this simplification.
In this article we reconsider the assumption of non-collisionality for galaxies in merging galaxy clusters. For this purpose, we extend the formalism developed by~\cite{Markevitch:2003at} to calculate the behaviour of a DM halo as a whole from the interactions of individual DM particles. Our calculations yield the evaporation and deceleration rates of a small (i.e.\ galaxy-sized) DM halo moving though a background DM density (a galaxy cluster) as a function of its escape velocity and its velocity relative to the background density. The resulting expressions provide an effective description of DM self-interactions, which can be used to capture the dominant effects of SIDM on DM haloes that are not explicitly included in a simulation.

In contrast to~\cite{Markevitch:2003at}, we pay special attention to the case in which the self-interaction cross section depends on the relative velocity of the DM particles and the scattering angle, and we discuss the case of long-range interactions as a specific example~\citep{Ackerman:2008gi,Feng:2009mn,Tulin:2013teo,Kaplinghat:2015aga,Agrawal:2016quu}. We demonstrate that the effects of SIDM cannot be captured by the momentum transfer cross section alone, but that 
two independent quantities are necessary to represent the underlying particle physics: the \emph{evaporation fraction} $\chi_\mathrm{e}$ and the \emph{deceleration fraction} $\chi_\mathrm{d}$. 

To illustrate our results with a well-motivated example, we study the behaviour of galaxies in a merging galaxy cluster and compare the results obtained with our effective description to the case of collisionless test particles. We show that DM self-interactions can significantly alter the trajectories of the galaxies and therefore reduce the predicted offset between the galaxies and the DM halo. These effects are particularly pronounced in the case of self-interactions with a strong angular dependence, for which evaporation is suppressed, allowing for a larger deceleration.

This paper is organised as follows. In Sec.~\ref{sec:scattering} we derive the deceleration and evaporation rates of a DM halo from the fundamental interactions of individual DM particles with a special emphasis on anisotropic scattering. We provide a system of differential equations that effectively describes the evolution of small DM haloes in the presence of DM self-interactions. Section~\ref{sec:mergers} focusses on the implementation of these equations to the specific case of galaxies in merging galaxy clusters. A simple numerical simulation is introduced in Sec.~\ref{sec:simulations} and the results are presented. Section~\ref{sec:discussion} provides a discussion of these results and our conclusions. Additional details are provided in the appendices~\ref{app:assumptions} and~\ref{app:1deddington}.

\section{Deceleration and evaporation}
\label{sec:scattering}

In this section we discuss how the scattering of individual DM particles gives rise to the evaporation and deceleration of a DM halo moving through a background DM density, such as a galaxy moving through a galaxy cluster. This approach follows closely the one from~\cite{Markevitch:2003at}, subsequently applied by~\cite{Kahlhoefer:2013dca} and~\cite{Kim:2016ujt}, except that we do not limit ourselves to the case of isotropic scattering but instead consider arbitrary DM self-scattering cross sections $\mathrm{d}\sigma / \mathrm{d} \theta$. We begin by discussing the kinematics of a single scattering event and then derive the macroscopic effects resulting from a large number of scatters.

\subsection{Scattering of individual particles}

\begin{figure}
\begin{center}
\includegraphics[width=0.95\columnwidth]{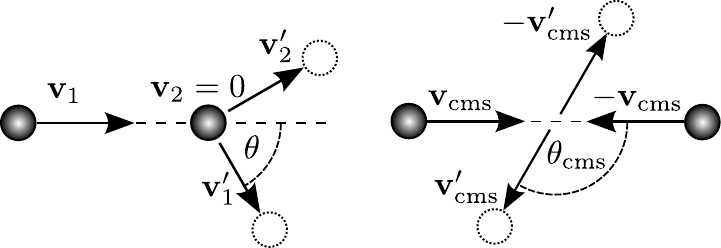}
\vspace{1mm}
\caption{\label{fig:variables}Collision between two DM particles in the rest frame of one of the particles (left) and in the centre-of-mass frame (right).}
\end{center}
\end{figure}

For the purpose of describing a single scattering process on the particle level, it is most convenient to consider the frame in which the DM halo is at rest (see left panel of Fig.~\ref{fig:variables}). We assume the relative velocity $v_0$ between the DM halo and the background density to be sufficiently large compared to the velocity dispersion of the background DM particles, such that all incoming DM particles approximately have a velocity of $\mathbf{v}_0=(0,0,v_0)$.\footnote{We carefully review this and the following assumptions in App.~\ref{app:assumptions}.} Similarly, we neglect the velocity dispersion of the DM particles bound to the halo, i.e.\ we approximate their velocity by $\mathbf{v}_2=(0,0,0)$. The infalling DM particles accelerate in the gravitational potential of the DM halo, so when they reach the central region, their velocity is $\mathbf{v}_1=\left(0,0,v\right)$, where $v = \sqrt{v_0^2+v_\text{esc}^2}$ and $v_{\text{esc}}$ is the local escape velocity at the point of scattering. 

To make contact to the angular distribution of the scattering cross section, we evaluate the scattering process in the centre-of-mass system (cms) and then transform back to the rest frame of the halo (see right panel of Fig.~\ref{fig:variables}).\footnote{Note that $\theta_\text{cms}$ is related to the scattering angle in the rest frame of the halo by $\theta_\text{cms} = 2 \theta$.} In terms of the scattering angle $\theta_\text{cms}$ in the cms frame, the post-scattering velocities are
\begin{align}
\mathbf{v}^{\prime}_{1}&=\begin{pmatrix}0\\ \frac{1}{2} v \sin{\theta_{\text{cms}}}\\ \frac{1}{2} v (1+\cos{\theta_{\text{cms}}})\end{pmatrix}, \ \
\mathbf{v}^{\prime}_{2}=\begin{pmatrix}0\\-\frac{1}{2} v \sin{\theta_{\text{cms}}}\\ \frac{1}{2} v (1-\cos{\theta_{\text{cms}}})\end{pmatrix},
\end{align}
where we have chosen the $x$-axis perpendicular to the scattering plane. If the magnitude of either of these velocities is larger than the local escape velocity of the halo, the corresponding particle is no longer bound to the halo and will escape. For particle 1 this will be the case if 
\begin{equation}
  \theta_{\text{cms}} < \theta_\text{crit} \equiv \arccos\left(\frac{x^2-1}{1+x^2}\right) \; ,
\end{equation}
where we have defined $x = v_\text{esc} / v_0$. Correspondingly, the second particle escapes if $\theta_\text{cms} > \pi - \theta_\text{crit}$. If $x < 1$ and hence $\theta_\text{crit} > \pi/2$, it is possible for both particles to escape from the halo, provided $\pi - \theta_\text{crit} < \theta_\text{cms} < \theta_\text{crit}$. In this case the halo effectively loses particles and evaporates. The probability for this to happen can be calculated in terms of the differential scattering cross section:
\begin{equation}
\chi_{\text{e}} = \frac{1}{\sigma}\int^{\theta_{\text{crit}}}_{\pi-\theta_{\text{crit}}}\frac{\text{d}\sigma}{\text{d}\theta}\text{d}\theta = \frac{1}{\sigma}\int^{\cos \theta_{\text{crit}}}_{- \cos \theta_{\text{crit}}}\frac{\text{d}\sigma}{\text{d}\cos \theta} \sin \theta \, \text{d}\cos \theta \; , \label{eq:chie}
\end{equation}  
where $\sigma = \int\frac{\text{d}\sigma}{\text{d}\theta}\text{d}\theta$. We refer to $\chi_\text{e}$ as the \emph{evaporation fraction}. In the case of isotropic scattering, where \mbox{$\text{d}\sigma/\text{d}\theta = \tfrac{1}{2} \, \sigma \sin\theta$}, we obtain~\citep{Markevitch:2003at}
\begin{equation}
\chi_{\text{e}}=\frac{1-x^2}{1+x^2} \; .
\end{equation}
Note that $\chi_\text{e}$ becomes negative if $v_0 < v_\text{esc}$, in which case it is impossible for both particles to escape but instead the DM halo can capture the incoming DM particle. In other words, rather than evaporating, on average the halo acquires additional mass during a scattering process.

In addition to evaporating the DM halo will also decelerate. The reason is that after the scattering process the escaping DM particles decelerate in the gravitational field of the DM halo. The sum of the momenta of the escaping DM particles will then in general be smaller than the momentum of the original infalling DM particle. In other words, the gravitational attraction between the escaping DM particles and the DM halo causes the DM halo to acquire some of their momentum. 

After having left the halo, the particles will have the velocities $v_{i,\text{E}}=\sqrt{|\mathbf{v}^{\prime}_{i}|^2-v_{\text{esc}}^2}$. If the interaction point is not too far from the centre of the DM halo, we can assume that the particles move approximately on radial orbits so that $\mathbf{v}_{i,\text{E}}$ points in the same direction as $\mathbf{v}^{\prime}_{i}$.\footnote{As we discuss in Appendix~\ref{app:assumptions}, the average scattering position is close to the scale radius of the halo. Nevertheless, the largest effects arise from the most tightly bound particles, which are found close to the centre of the system.} We then find
\begin{align}
\mathbf{v}_{1,\text{E}}&=\begin{pmatrix}0\\\frac{v_0}{\sqrt{2}}\sin\frac{\theta_{\text{cms}}}{2} \sqrt{1-x^2+ (1+x^2)\cos{\theta_{\text{cms}}}}\\\frac{v_0}{\sqrt{2}}\cos\frac{\theta_{\text{cms}}}{2}\sqrt{1-x^2+(1+x^2)\cos{\theta_{\text{cms}}}}\end{pmatrix}\;,\label{eq:v1E}\\
\mathbf{v}_{2,\text{E}}&=\begin{pmatrix}0\\-\frac{v_0}{\sqrt{2}}\cos\frac{\theta_{\text{cms}}}{2}\sqrt{1-x^2-(1+x^2) \cos{\theta_{\text{cms}}}}\\\frac{v_0}{\sqrt{2}}\sin\frac{\theta_{\text{cms}}}{2}\sqrt{1-x^2-(1+x^2) \cos{\theta_{\text{cms}}}}\end{pmatrix}\;.\label{eq:v2E}
\end{align}
It is understood implicitly that $\mathbf{v}_{1,\text{E}} = 0$ for $\theta_\text{cms} > \theta_\text{crit}$ and $\mathbf{v}_{2,\text{E}} = 0$ for $\theta_\text{cms} < \pi - \theta_\text{crit}$.

The total change in momentum of the halo can be calculated by comparing the incoming momentum with the total outgoing momentum:
\begin{align}
\Delta \mathbf{p}&=m_{\chi}(\mathbf{v}_0-\mathbf{v}_{1,\text{E}}-\mathbf{v}_{2,\text{E}}) \; ,
\end{align} 
where $m_\chi$ denotes the DM mass. Although the net effect in the $y$-direction can be non-zero for a single scattering process, the effects will average to zero when considering a larger number of scatters. In the $z$-direction, on the other hand, a non-zero effect remains. We can therefore define the \emph{deceleration fraction} as
\begin{align}
\chi_d=\frac{1}{m_{\chi}\,v_0\,\sigma}\int \Delta p_{\text{z}}\frac{\text{d}\sigma}{\text{d}\theta}\text{d}\theta \; .
\end{align}

Substituting eqs.~(\ref{eq:v1E}) and (\ref{eq:v2E}), we obtain
\begin{align}
\chi_d=1-\frac{1}{\sqrt{2}\sigma}\int_0^{\theta_{\text{crit}}} & \cos\frac{\theta}{2}\sqrt{1-x^2+(1+x^2)\cos\theta} \nonumber \\
&\cdot\left(\frac{\text{d}\sigma}{\text{d}\theta}(\theta)+\frac{\text{d}\sigma}{\text{d}\theta}(\pi-\theta)\right)\text{d}\theta \; . \label{eq:chid}
\end{align}
Note that this expression correctly accounts also for the cases where one or both of the DM particles remain bound to the halo. If the cross section is symmetric under the replacement $\theta \to \pi - \theta$ this expression simplifies to
\begin{align}
\chi_d=1-\frac{\sqrt{2}}{\sigma}\int_0^{\theta_{\text{crit}}}\cos\frac{\theta}{2}\sqrt{1-x^2+(1+x^2)\cos\theta} \, \frac{\text{d}\sigma}{\text{d}\theta} \, \text{d}\theta.
\end{align}
From this expression it is straight-forward to recover the result for the isotropic case \citep{Markevitch:2003at}:
\begin{align}
\chi_d=1-4\int_{\sqrt{x^2 / (1 + x^2)}}^1y^2\sqrt{y^2-x^2(1-y^2)}\text{d}y \;.
\end{align}

\subsection{Anisotropic scattering}

For isotropic scattering the relative magnitude of $\chi_d$ and $\chi_e$ depends only on the value of $x = v_\text{esc} / v_0$. For small values of $x$ the evaporation fraction will be large, but the escaping particles are typically too fast to lead to a sizeable deceleration of the halo. For larger $x$, on the other hand, the majority of collisions will not change the number of particles bound to the DM halo and a significant fraction of momentum can be transferred.

\begin{figure}
\centering
\includegraphics[width = 0.79\columnwidth,clip,trim = 0 0 -5 0]{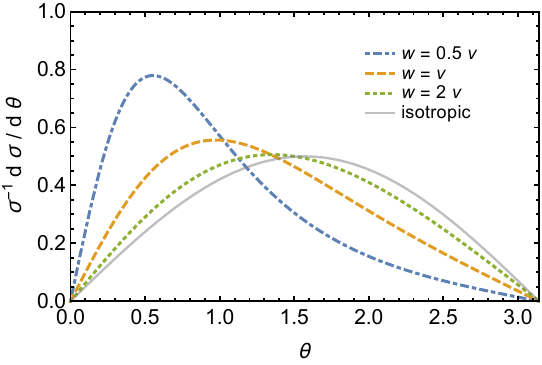}\\
\vspace{2mm}
\includegraphics[width = 0.8\columnwidth]{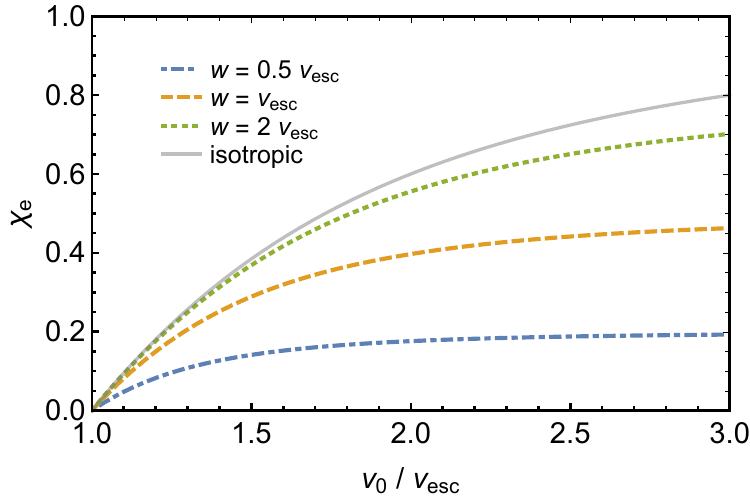}\\
\vspace{2mm}
\includegraphics[width = 0.8\columnwidth]{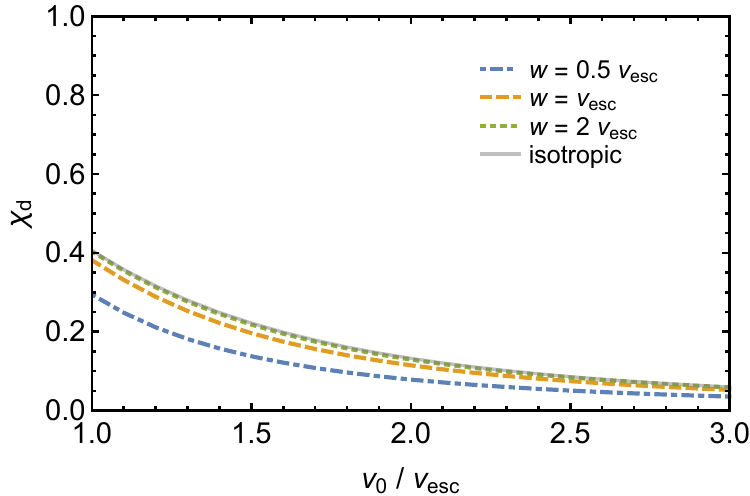}
\caption{The top panel shows the normalised differential scattering cross section for anisotropic scattering as a function of the scattering angle $\theta$ for different ratios of $v / w$. The centre and bottom panels show respectively the evaporation and deceleration fraction as a function of the velocity $v$ for different values of the cut-off velocity $w$. While the evaporation fraction depends sensitively on $w$, the deceleration fraction is nearly unaffected.
\label{fig:aniso}}
\end{figure}

The situation changes significantly if $\mathrm{d}\sigma / \mathrm{d}\theta$ exhibits a strong angular dependence~\citep{Kahlhoefer:2013dca,Robertson:2016qef}. For example, if the differential scattering cross section peaks at small scattering angles, it is possible that the evaporation fraction remains small even for small $x$, because scattering angles in the range $\pi - \theta_\text{crit} < \theta_\text{cms} < \theta_\text{crit}$ are less likely. A typical example often encountered in the context of SIDM is the case that DM particles interact with each other via the exchange of a light mediator with mass $m_\phi \lesssim m_\chi \, v_0$~\citep{Ackerman:2008gi,Feng:2009mn,Tulin:2013teo,Kaplinghat:2015aga,Agrawal:2016quu}. Focussing on the case of $t$-channel scattering\footnote{We note that in realistic models of SIDM $t$-channel scattering does not usually appear in isolation. For the scattering of two identical particles, also $u$-channel scattering will be present and restores the symmetry under particle exchange ($\theta \to \pi - \theta$). For the scattering of particle and antiparticle, on the other hand, $s$-channel scattering will also be possible, but usually only gives a negligible contribution. Our conclusions remain unchanged when including these additional scattering channels.}, the differential cross section is given by (see e.g.~\cite{Tulin:2017ara})
\begin{align}
\frac{\text{d}\sigma}{\text{d}\theta}&=\frac{2\pi\alpha^2m_{\chi}^2 \sin \theta}{\left(\frac{1}{2} m_{\chi}^2v^2 (1-\cos\theta)+m_{\phi}^2\right)^2}  \equiv \frac{\sigma_0 \sin \theta}{2\left(1+\frac{v^2}{w^2}\sin^2\frac{\theta}{2}\right)^2} \; ,\label{eq:longrange}
\end{align}
where $w=m_{\phi}/m_{\chi}$ and $\sigma_0=(4\pi\alpha^2m_{\chi}^2)/m_{\phi}^4$. For $w \ll v$ this leads to the well-known expression for long-range (Rutherford) scattering, while for $w \gg v$ the scattering cross section is approximately isotropic (see top panel in Fig.~\ref{fig:aniso}).

Long-range interactions have the crucial property that scattering with large momentum transfer is suppressed. As a result, we expect the evaporation fraction (which is most sensitive to large scattering angles) to depend sensitively on the velocity scale $w$. This expectation is confirmed in the middle and bottom panels of Fig.~\ref{fig:aniso}, which show the evaporation and deceleration fractions as a function of the velocity $v_0$ (in units of the escape velocity $v_\text{esc}$) for different values of the cut-off velocity $w$. We conclude that for different angular shapes of the scattering cross section the relative importance of evaporation and deceleration can change. We will now turn to the calculation of evaporation and deceleration \emph{rates} to make this statement more precise.

\subsection{Scattering rates}
\label{sec:rates}

Having derived the probability for evaporation and the average fraction of momentum lost in a single scattering process, we can now calculate the evaporation and deceleration rates arising from a large number of scattering events. The total scattering rate of a given DM particle in the halo is
\begin{equation}
R = \sigma \, v_0 \frac{\rho}{m_{\chi}} \; ,
\end{equation}
where $\rho$ denotes the mass density of the DM background. The \emph{evaporation rate} and \emph{deceleration rate} are then given by
\begin{equation}
 f_e=\chi_e(v_0,v_{\text{esc}}) \cdot R \; , \qquad f_d=\chi_d(v_0,v_{\text{esc}}) \cdot R \; .
\end{equation}
The average relative change in mass and velocity per time can then be written as
\begin{align}
\dot{\mathbf{v}} = -\mathbf{v} \cdot f_d \; , \qquad \dot{M} = -M \cdot f_e \; .
\end{align}

These equations can be used to obtain an effective description of small DM haloes, which are not explicitly resolved in a given simulation, in the presence of DM self-interactions. As discussed in the introduction, an interesting application is the simulation of galaxies in merging galaxy clusters (see Sec.~\ref{sec:mergers}). These galaxies are often simply represented by collisionless test particles even though each galaxy consists of both stars and DM and should therefore feel the effect of DM self-interactions. To implement this effect, let us assume that each test particle representing a galaxy is characterised not only by its position $\mathbf{x}$ and velocity $\mathbf{v}$, but also by its total mass in stars $M_\text{star}$, its total mass in DM $M_\text{DM}$ and its typical size $a$. The latter can be thought of for example as the scale radius of the Hernquist profile:
\begin{align}
\rho(M,a,r) = \frac{M}{2\pi}\frac{a}{r(r+a)^3} \; ,
\end{align}
where $M = M_\text{star} + M_\text{DM}$. It is important to note that the assumed values for $M_\text{star}$ and $M_\text{DM}$ need not be related to the mass of the test particles in the simulation. In other words, the masses $M_\text{star}$ and $M_\text{DM}$ should correspond to realistic galaxies and should \emph{not} depend on the number of test particles in the simulation. These masses are used first of all to calculate the escape velocity of the galaxy. For a Hernquist profile, the escape velocity is given by
\begin{align}
v_{\text{esc}}(M,a,r)=\sqrt{\frac{2 \, G \, M}{r+a}} \; ,
\end{align}
where $G$ denotes Newton's constant. Of course, $v_\text{esc}$ depends on the radius $r$, but we can approximately assume that a typical scattering event occurs at $r \approx a$ (this approximation is confirmed by explicit numerical simulation in App.~\ref{app:assumptions}). The average escape velocity is hence simply $\overline{v}_\text{esc} = \sqrt{G M / a}$.

The equations of motion for a galaxy test particle moving in the gravitational potential $\Phi(\mathbf{x})$ created by a DM distribution $\rho(\mathbf{x})$ are then given by
\begin{align}
& \dot{\mathbf{v}} = -\nabla\Phi(\mathbf{x}) - \frac{M_\text{DM}}{M_\text{DM} + M_\text{star}} \mathbf{v} \, \chi_d(v_0, v_\text{esc}) \frac{\sigma \, v_0 \, \rho(\mathbf{x})}{m_{\chi}} \; , \label{eq:decel}\\
& \dot{M}_\text{DM} = - M_{\text{DM}} \, \chi_e(v_0, v_{\text{esc}})\frac{\sigma \, v_0 \, \rho(\mathbf{x})}{m_{\chi}} \; \label{eq:evap},
\end{align}
where $v_0$ denotes the relative velocity between the test particle and the DM distribution.\footnote{In practice, $v_0$ and $\rho$ are calculated by averaging the velocity and density of simulated DM particles in the vicinity of the test particle. While the size of the volume included in this averaging may depend on the resolution of the simulation, it should not be smaller than the actual size of the test particle system, characterized by $a$.} For simplicity we assume that to first order DM self-interactions do not affect the size of the system ($\dot{a} = 0$) and do not lead to the evaporation of stars ($\dot{M}_\text{star} = 0$). In other words, the evaporation only reduces the DM part of the galaxy mass, but not the baryonic part. This reduction in  $M_{\text{DM}}$ then changes the escape velocity of the galaxy, which enters into the calculation of $\chi_\mathrm{e}$ and $\chi_\mathrm{d}$ as well as the ratio $M_{\text{star}}/M_{\text{DM}}$.

To give an illustration, let us consider the motion of a test particle representing a typical galaxy through a homogeneous DM density ($\Phi(\mathbf{x}) = 0$ and $\rho(\mathbf{x}) = \rho_0$). In order to compare the evolution of $M_\text{DM}$ and $v$ as a function of time for cross sections with different angular dependence, we evaluate the momentum transfer cross section \citep{Robertson:2016qef,Kahlhoefer:2017umn}\footnote{Often the momentum transfer cross section is defined as $ \sigma_\mathrm{T} \equiv \int \frac{\mathrm{d}\sigma}{\mathrm{d}\theta} (1 - \cos \theta) \mathrm{d}\theta$ \citep{Tulin:2013teo,Agrawal:2016quu}. This definition however is not suitable for indistinguishable particles \citep{Kahlhoefer:2017umn} and does not correctly predict the rate of core formation in DM haloes~\citep{Robertson:2016qef}.}
\begin{equation}
 \sigma_\mathrm{T} \equiv \int \frac{\mathrm{d}\sigma}{\mathrm{d}\theta} (1 - |\cos \theta|) \mathrm{d}\theta
\end{equation}
and require that all cross sections under consideration correspond to the same momentum transfer cross section at a given reference velocity $v_\text{ref}$.

\begin{figure}
\centering
\includegraphics[width = 0.8\columnwidth]{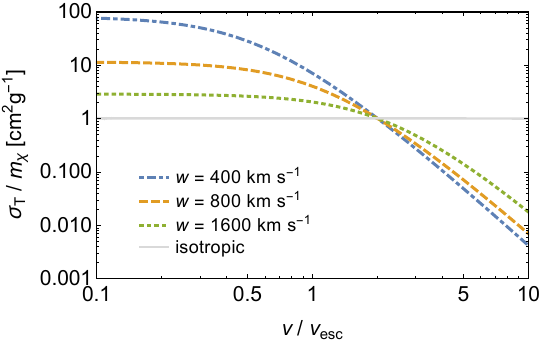}\\
\vspace{2mm}
\includegraphics[width = 0.81\columnwidth,clip,trim = -8 0 0 0]{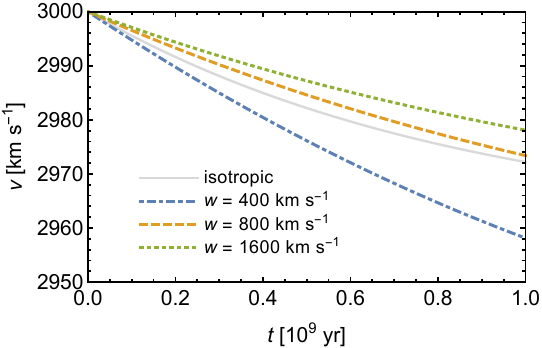}\\
\vspace{2mm}
\includegraphics[width = 0.81\columnwidth,clip,trim = -6 0 0 0]{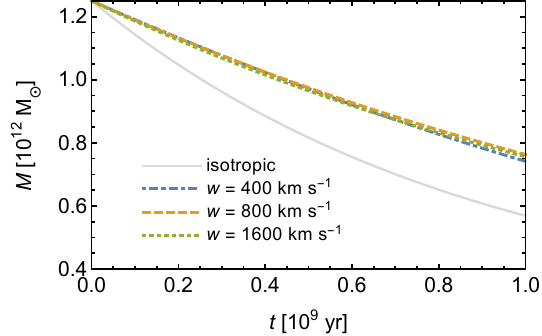}\\
\caption{The top panel shows the momentum transfer cross section for isotropic and anisotropic scattering (parametrized by the cut-off velocity $w$), scaled to satisfy $\sigma_\mathrm{T}(v_\text{ref} = 2000 \:\mathrm{km \, s^{-1}}) / m_\chi = 1\:\mathrm{cm^2 \, g^{-1}}$. The centre and bottom panel show respectively the evolution of the velocity and the total mass of a galaxy-sized halo moving through a uniform DM background. See text for details on the assumed astrophysical parameters.\label{fig:testparticle}}
\end{figure}

For concreteness, we take $M_\text{DM,0} = 10^{12} \text{M}_{\odot}$, $M_\text{star,0} = 2.5 \cdot 10^{11} \text{M}_{\odot}$ and $a = 20\,\text{kpc}$, such that $v_\text{esc,0} \approx 530 \, \mathrm{km\,s^{-1}}$, as well as $v_0 = 3000 \, \mathrm{km\,s^{-1}}$ and $\rho_0 = 10^{-3} \text{M}_{\odot} \mathrm{pc^{-3}}$. We consider the anisotropic scattering cross section from long-range interactions introduced in eq.~(\ref{eq:longrange}) for different values of $w$ and fix the normalization $\sigma_0$ by the requirement $\sigma_\mathrm{T}(v_\text{ref} = 2000\,\mathrm{km\,s^{-1}})/m_\chi = 1 \mathrm{cm^2 g^{-1}}$. Our results are shown in Fig.~\ref{fig:testparticle}. As expected, we see that small values of $w$ lead to a differential cross section that favours deceleration relative to evaporation. Correspondingly, the velocity decreases most rapidly if $w$ is small, whereas the decrease of the total mass of the galaxy is significantly smaller compared to the isotropic case, but approximately of the same size for all $w$. 

These observations obviously depend to some degree on the choice of the matching scale. Lower values of $v_\text{ref}$ lead to a further suppression of the effects for anisotropic scattering relative to the isotropic case. This dependence on $v_\text{ref}$ is strongest for the cross section with the strongest angular dependence (i.e.\ with the smallest $w$). For Fig.~\ref{fig:testparticle} we have chosen $v_\text{ref}$ in such a way that the evaporation rate is similar for all choices of $w$ and the deceleration rates differ significantly. Reducing $v_\text{ref}$ would lead to more similar deceleration rates but a larger disparity in the evaporation rates.\footnote{Note that increasing $v_\text{ref}$ beyond $2000\,\mathrm{km\,s^{-1}}$, or lowering $w$ below $400\,\mathrm{km\,s^{-1}}$, would lead to unacceptably large values of $\sigma_\mathrm{T} / m_\chi$ at small scales~\citep{Tulin:2017ara}.}

To conclude this section, let us perform an important consistency check. Our effective description is based on the assumption that DM particles are very unlikely to scatter more than once on the relevant dynamical time scales. In particular, we have assumed that if both DM particles have a sufficiently large velocity after the collision to escape from the galaxy, they will do so without scattering again. We can check that this assumption is valid for the case considered above by calculating the surface density of the galactic halo for the adopted parameters. We find that the average surface density within $r < a$ is of order $0.05 \, \mathrm{g \, cm^{-2}}$, such that the probability to scatter again is small as long as $\sigma / m_\chi \ll 20 \, \mathrm{cm^2 \, g^{-1}}$.

\section{Galaxies in merging clusters}
\label{sec:mergers}

Having derived the effective equations that describe the motion of small DM haloes, we now turn to the central application of these equations: the simulation of the motion of galaxies in merging galaxy clusters. Major mergers are one of the most promising probes of DM self-interactions, as they may be sensitive not only to the total self-interaction rate but also to the angular distribution of the DM self-scattering~\citep{Kahlhoefer:2013dca}. A useful observable in this context is the separation between the centroids of the distribution of galaxies and the DM distribution inferred from gravitational lensing. The basic argument is that in the absence of DM self-interactions galaxies and DM particles are both effectively collisionless and will therefore behave in the same way. Any difference in their distributions after a cluster collision would therefore point to SIDM.

As we have discussed in detail above, however, in the presence of DM self-interactions galaxies also no longer behave as perfectly collisionless particles. Numerical simulations treating galaxies as collisionless therefore risk overestimating the differences in the behaviour of DM and galaxies and may therefore find larger centroid separations than expected in a more realistic treatment. In the remainder of this work we investigate whether the effect of DM self-interactions on galaxies can significantly modify the predicted separations obtained from numerical simulations. We will focus on an unequal merger like the Bullet Cluster and develop a simplified description to estimate the magnitude of the expected effects.

\begin{figure}
\centering
\includegraphics[width = 0.95\columnwidth]{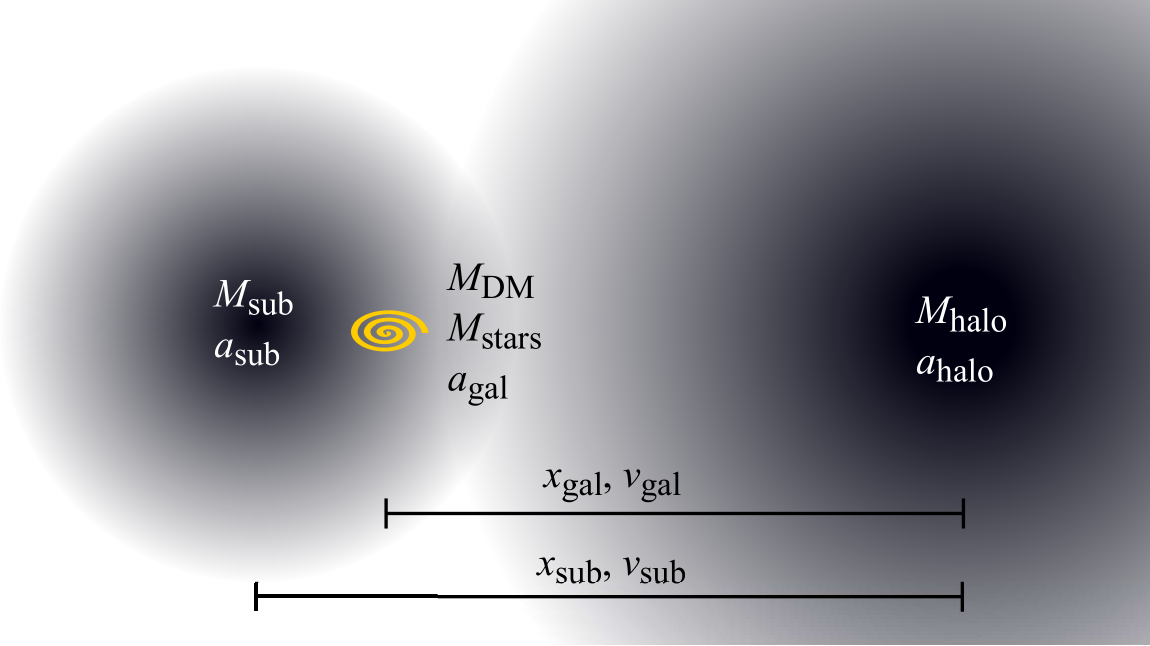}\\
\caption{Sketch of the two merging galaxy clusters and the various quantities defined in the text.\label{fig:sketch}}
\end{figure}

For this purpose we consider a system of two galaxy clusters (see illustration in Fig.~\ref{fig:sketch}). The main cluster is modelled by a Hernquist profile with mass $M_{\text{halo}}$ and scale radius $a_{\text{halo}}$ and is assumed to be at rest at $\mathbf{x}=0$. The smaller cluster (subsequently called the sub-cluster) has mass $M_{\text{sub}}$ and scale radius $a_{\text{sub}}$. Its velocity $\mathbf{v}_{\text{sub}}$ and position $\mathbf{x}_{\text{sub}}$ will vary with time as it falls into the main cluster. The advantage of considering a system with two unequal masses is that we can~-- to first approximation~-- make use of the effective equations derived above to model the effects of DM self-interactions on the motion of the sub-cluster, i.e.\ we can calculate the evaporation and deceleration rates as a function of the DM density of the main cluster and the relative velocity between the two haloes:
\begin{align}
\dot{\mathbf{v}}_{\text{sub}} = & -\nabla\Phi_{\text{halo}}(|\mathbf{x}_{\text{sub}}|) \nonumber \\
& -\mathbf{v}_{\text{sub}} \, \chi_d(v_{\text{sub}},v_{\text{esc,sub}}) \, \sigma \, v_{\text{sub}} \frac{\rho_{\text{halo}}(|\mathbf{x}_{\text{sub}}|)}{m_{\chi}} \; , \label{eq:vsub}\\
\dot{M}_{\text{sub}} = & -M_{\text{sub}} \, \chi_e(v_{\text{sub}},v_{\text{esc,sub}}) \, \sigma \, v_{\text{sub}} \frac{\rho_{\text{halo}}(|\mathbf{x}_{\text{sub}}|)}{m_{\chi}} \; , \label{eq:Msub}
\end{align}
where $v_\text{esc,sub}$ depends implicitly on $M_\text{sub}$ and $a_\text{sub}$. Note that here we do not make a distinction between the total mass of the sub-cluster $M_\text{sub,tot}$ and the mass of its DM halo $M_\text{sub}$, i.e.\ we assume that $M_\text{sub,tot} \approx M_\text{sub}$. In particular, we do not model the effects of intracluster gas, which gives a sub-dominant contribution to the gravitational potential of the sub-cluster, and only consider the contribution of DM and galaxies (see below). While this approach is too crude to derive reliable estimates of the DM-galaxy separation,\footnote{In particular, we do not include the deformation of the sub-cluster due to the tail of escaping DM particles, which has been shown to significantly shift the centroid of the DM halo \citep{Kahlhoefer:2013dca}.} it is fully sufficient 
to study the effect of DM self-interactions on the motion of individual galaxies.

An obvious problem with eqs.~(\ref{eq:vsub}) and (\ref{eq:Msub}) is that for a Hernquist profile $\rho_\text{halo}(r)$ diverges for $r \to 0$, introducing an unphysical dependence on the precise value of the impact parameter. This unphysical behaviour can be avoided by averaging the density profile over a sphere with radius given by the size of the sub-cluster, i.e.\ we replace $\rho_\text{halo}$ by
\begin{align}
\rho_{\text{avg}}(r,a)=\frac{3}{2 \, a^3}\int^1_{-1} & \text{d}\cos\theta \int_0^{a} \text{d}r^{\prime} r^{\prime 2} \nonumber \\ & \cdot \rho(\sqrt{r^2+2rr^{\prime}\cos\theta+r^{\prime 2}}).
\end{align}
We also note that in contrast to the simplified discussion in Sec.~\ref{sec:scattering} we can no longer neglect the velocity dispersion $v_\text{disp}$ of the DM particles in the main cluster. As discussed in detail in App.~\ref{app:assumptions}, this additional contribution can be adequately taken into account by replacing 
\begin{align}
\chi_e(v, v_\text{esc}) \quad \rightarrow  \quad \chi_e \left(\sqrt{v^2+v_{\text{disp}}^2}, v_{\text{esc}}\right) \;.
\end{align}
For the deceleration rate the effect of a non-zero velocity dispersion is more complicated since not only the magnitude of the velocities of the incoming DM particles matters but also the direction. To avoid extrapolation into an unphysical regime, we include deceleration only if $v>v_{\text{disp}}$ and set $\chi_d$ to zero otherwise. Since $v_\text{sub} \gg v_\text{disp}$ at core passage, this modification has a very minor effect.

Once we have calculated the trajectory of the infalling sub-cluster in the presence of DM self-interactions, we can study the motion of galaxies initially bound to this sub-cluster. As before, the total mass of each galaxy is given by the sum of the DM component and the stellar component, $M_\text{gal} = M_\text{star} + M_\text{DM}$ and we assume that $M_\text{star}$ does not change. The underlying differential equations are given by
\begin{align}
\dot{\mathbf{v}}_{\text{gal}}= & -\nabla \Phi_{\text{halo}}(|\mathbf{x}_{\text{gal}}|) -\nabla \Phi_{\text{sub}}(|\mathbf{x}_{\text{gal}}-\mathbf{x}_{\text{sub}}|) \nonumber \\
&-\frac{\mathbf{v}_{\text{gal}} \, \chi_d(v_{\text{gal}},v_{\text{esc,gal}}) \, M_\text{DM}}{M_\text{DM} + M_\text{star}}  \, \sigma \, v_{\text{gal}} \, \frac{\rho_{\text{avg}}(|\mathbf{x}_{\text{gal}}|,a_\text{gal})}{m_{\chi}} \nonumber \\ 
&-\mathbf{v}_{\text{sub}} \, \chi_d(v_{\text{sub}},v_{\text{esc,sub}}) \, \sigma \, v_{\text{sub}} \, \frac{\rho_{\text{avg}}(|\mathbf{x}_{\text{sub}}|,a_\text{sub})}{m_{\chi}} \; , \label{eq:vgal}\\
\dot{M}_{\text{DM}} = & -M_{\text{DM}} \, \chi_e(v_{\text{gal}},v_{\text{esc,gal}}) \, \sigma \, v_{\text{gal}} \, \frac{\rho_{\text{avg}}(|\mathbf{x}_{\text{gal}}|,a_\text{gal})}{m_{\chi}} \; . \label{eq:Mgal}
\end{align}
A number of comments are in order: First, the final term in eq.~(\ref{eq:vgal}) is identical to the final term in eq.~(\ref{eq:vsub}). The reason that this term appears in the differential equation for $v_\text{gal}$ is that the deceleration of the sub-cluster is a gravitational effect, caused by the capture of DM particles as well as by the tail of particles escaping from the sub-cluster. This gravitational attraction is equally felt by the galaxies bound to the sub-cluster and causes the galaxies to decelerate as well. Note that this term would be present even if galaxies were perfectly collisionless (e.g.\ because $M_\text{DM} = 0$). In addition the galaxies feel a deceleration effect due to scattering processes of DM particles bound to the galactic halo with the background of the main cluster. As a result, the galaxies actually experience a larger deceleration rate than the sub-cluster and~-- in the absence of any other effect~-- would therefore be expected to \emph{lag behind} the DM halo. This effect is however expected to be overcompensated by the shift of the DM centroid due to the deformation of the DM halo, which is not included in our simulation.

The second important comment is that eqs.~(\ref{eq:vgal}) and~(\ref{eq:Mgal}) do \emph{not} contain terms to model the effect of interactions between DM particles from the galaxies and DM particles from the sub-cluster, i.e.\ there is no term proportional to
$ \chi_\text{d,e}(|\mathbf{v}_\text{gal} - \mathbf{v}_\text{sub}|, v_\text{esc,gal}) \, |\mathbf{v}_\text{gal} - \mathbf{v}_\text{sub}| \, \rho_\text{sub}(|\mathbf{x}_\text{gal} - \mathbf{x}_\text{sub}|) $. The reason is that for a typical galaxy bound to the sub-cluster one finds $|\mathbf{v}_\text{gal} - \mathbf{v}_\text{sub}| \sim v_\text{disp,sub} \sim v_\text{esc,gal}$. In this case our effective description becomes less reliable, but we expect the evaporation rate to be rather small due to the non-negligible probability of particle capture and the deceleration rate to be small because the effect of different interactions average out (see App.~\ref{app:assumptions}). Moreover, the effect of the sub-cluster on the galaxies is additionally suppressed because $\sigma | \mathbf{v}_\text{gal} - \mathbf{v}_\text{sub}| \rho_\text{sub}$ is small compared to $\sigma |\mathbf{v}_\text{gal}| \rho_\text{halo}$. This suppression can however be compensated if the cross sections has a strong velocity dependence. We will return to this issue in Sec.~\ref{sec:discussion}.

\section{Numerical simulation}
\label{sec:simulations}

A number of different groups have performed numerical simulations of the Bullet Cluster and have used the results to obtain constraints on DM self-interactions. One of the central results is that for cross sections $\sigma_\mathrm{T} / m_\chi \sim 1 \mathrm{cm^2 \, g^{-1}}$ one can expect a separation between the centroids of DM and galaxies in the sub-cluster in the range from $\sim 10\,\text{kpc}$ (\cite{Robertson:2016xjh}) to $\sim 40\,\text{kpc}$ (\cite{Randall:2007ph}), which is about a factor of 2 below current observational bounds. The question we want to address here is whether these simulations may actually overestimate the DM-galaxy separation by assuming galaxies to be collisionless test particles. 

For this purpose we have implemented the differential equations described above in a numerical simulation. Due to the simplified nature of this simulation, we will not attempt to map our results onto observations to extract predictions for the DM-galaxy separation. Instead, our main objective is to estimate the effect of DM self-interactions on the centroid of the galaxy distribution, i.e.\ we will compare the position of the galaxy distribution obtained when including our effective description of DM self-interactions to the case where the galaxies are assumed to be collisionless test particles. In this section we describe the assumed astrophysical parameters and the set-up of the simulation and then present our results.

\subsection{Astrophysical parameters}

The Bullet Cluster is a well-studied example for the kind of system we are interested in~\citep{Markevitch:2003at,Clowe:2006eq,Randall:2007ph,Lage:2013yxa,Robertson:2016xjh}. While we do not attempt to model the Bullet Cluster realistically, we use it as an orientation in order to determine reasonable astrophysical parameters for the various components of the system we simulate. For the main cluster we adopt the values $M_{\text{halo}}=3.85\cdot 10^{15}  \text{M}_{\odot}$ and $a _{\text{halo}}=1290\ \text{kpc}$, while we model the sub-cluster with $M_{\text{sub,tot}}=2.5\cdot 10^{14}  \text{M}_{\odot}$ and $\ a _{\text{sub}}=280\ \text{kpc}$ (using the values suggested by \cite{Robertson:2016xjh}). 

The total mass of the sub-cluster can be written as $M_{\text{sub,tot}}=M_{\text{sub}}+N\cdot M_{\text{gal}}$, where $N$ is the total number of galactic sub-haloes bound to the sub-cluster and $M_{\text{sub}}$ is the mass of the smoothly distributed DM component. Following the study by~\cite{vandenBosch:2004zs} we assume that the sub-haloes make up around 5 per cent of the sub-cluster mass. These sub-haloes are not expected to contain any significant amount of gas, but they will in general contain stars, i.e.\ $M_{\text{gal}}=M_{\text{DM}}+M_{\text{star}}$. As pointed out by~\cite{Li:2015tps} the ratio of galaxy stellar mass to sub-halo mass is much larger for sub-haloes in the centre of a galaxy cluster than for field galaxies. Here we take this ratio to be $M_{\text{star}} / M_{\text{gal}} = 0.2$, such that $1\%$ of the total cluster mass is in stars, in agreement with the value suggested by~\cite{Robertson:2016xjh}, see also the study by~\cite{2012ApJ}.

From observations of the Bullet Cluster we know that the main cluster consists of 71 and the sub-cluster of 7 galaxies~\citep{Barrena:2002dp}. To account for the fact that some galaxies may have been tidally stripped from the sub-cluster during the core passage, we take an initial value of $N=10$. This choice implies $M_{\text{gal}} = 1.25\cdot 10^{12}\ \text{M}_{\odot}$ and hence $M_{\text{DM}}=1.0\cdot 10^{12}\  \text{M}_{\odot}$ and $M_{\text{star}}=2.5\cdot 10^{11}\ \text{M}_{\odot}$ for each of the galaxies. A typical scale radius of a galaxy is around $a_{\text{gal}}=20\ \text{kpc}$.

For the purpose of simulating the behaviour of the galaxies during the cluster collision we however do not have to limit ourselves to simulating only $N = 10$ galaxies. In fact, doing so would introduce large amounts of statistical noise and would make it very difficult to extract robust predictions independent of the specific initial conditions. The great advantage of having an effective description of the motion of galactic haloes is that we can simulate a much larger number of galaxies without the need for each simulated galaxy to be dynamically self-consistent. In other words, we can simply take the parameters derived above for $N = 10$ and use them to simulate $N = 1000$ galaxies. For simplicity we assume all galaxies to be described by the same parameters, although it would be straight-foward to consider a distribution of sub-halo masses, scale radii and stellar mass fractions. We will now describe how such a simulation is set up in detail.

\subsection{Setting up the simulation}

As discussed above, one of the great advantages of having an effective description of self-interactions is to enable the simulation of a large number of galaxies. The reason that large samples are important is that we expect the effect of DM self-interactions on galaxies to depend on their position and velocity relative to the sub-cluster during the core passage of the sub-cluster through the main cluster. Galaxies that are tightly bound and close to the centre of the sub-cluster are expected to be affected less strongly than the ones that are further away. Meaningful statements about the typical effect of DM self-interactions can therefore only be obtained by averaging over a sufficiently large sample size.

It is therefore essential to consider a realistic and self-consistent distribution of galaxies in the sub-cluster. Nevertheless, the shape of the individual orbits (i.e.\ whether the galaxies move on circular or radial orbits and how the orbits are oriented relative to the merger axis) is expected to be less important than the corresponding binding energies. We can therefore simplify the simulation by limiting ourselves to the one-dimensional case in which all galaxies travel on radial orbits aligned along the merger axis.\footnote{A limitation of the one-dimensional approach is that all galaxies pass directly through the central region of the main cluster (i.e. they have vanishing impact parameter). The effect of non-zero impact parameter can however be approximately captured by increasing the radius used for the density averaging. A comprehensive study of the effect of varying impact parameters (as well as different density profiles in the central region) is beyond the scope of the current work.}

To initialize the simulation we then need a random sample of galaxies from a self-consistent one-dimensional distribution function. Such a distribution function can be derived by making use of a modified version of the Eddington inversion formula, which allows to calculate a velocity distribution for given gravitational potential and density profile. As discussed in detail in App.~\ref{app:1deddington}, we take the galaxy density profile to be\footnote{We note that it is perfectly possible within this approach to assume that the density profile of galaxies is different from the one of DM, which gives the dominant contribution to the gravitational potential.} 
\begin{equation}
 \rho(x) \propto \frac{a}{(|x| + a)^2}
\end{equation}
and calculate the appropriate velocity distribution such that the distribution function is time-independent for an isolated sub-cluster.

The simulations discussed in the following are based on randomly generated samples of 1000 galaxies each. We have confirmed that this number is sufficiently large so that the results for different samples yield qualitatively identical conclusions. We will denote the ensemble-averaged position and velocity of the galaxies by $\overline{x_\text{gal}}$ and $\overline{v_\text{gal}}$, respectively. By construction, the initial conditions then satisfy $\overline{x_\text{gal,0}} = x_\text{sub,0}$ and $\overline{v_\text{gal,0}} = v_\text{sub,0}$. Over the course of the simulation the averaging can become significantly biased by galaxies that are no longer bound to the sub-cluster, for example because they have been tidally stripped by the main cluster. To avoid such systematic biases it is essential to restrict the averaging to galaxies that remain bound to the sub-cluster, for which we require $|x_{\text{sub}}-x_{\text{gal}}|<2a_{\text{sub}}$.

The initial position and velocity of the sub-cluster are taken to be $x_{\text{sub},0}=-1500\, \text{kpc}$ and $v_{\text{sub},0}=1000 \, \mathrm{km\,s^{-1}}$, respectively, such that we obtain a realistic collision velocity of approximately $v_{\text{coll}}\sim 3500 \, \mathrm{km\,s^{-1}}$.

\subsection{Results}

With the astrophysical parameters and the initial conditions discussed above, we can now proceed to solving the differential equations presented in Sec.~\ref{sec:mergers} in one dimension. As discussed above, the question that we will be interested in is whether it is a good approximation to treat galaxies as collisionless test particles or whether the effective description of DM self-interactions derived above leads to sizeable corrections. For this purpose, we run two simulations for each set of initial conditions: one without and one with the effect of SIDM on galaxies. In the first case the mass of each galaxy will remain unchanged over the course of the simulation, whereas in the latter case evaporation will occur: $\overline{M_\text{gal}} < \overline{M_\text{gal,0}}$.\footnote{Note that we do not include tidal effects on the galaxies, although our effective description could be extended to include them.} The halo-galaxy separation $\overline{\Delta x} = \overline{x_\text{gal}} - x_\text{sub}$ is more complicated, as this quantity can become non-zero even in the absence of self-interactions~\citep{Schaller:2015tia,Ng:2017bwr} and furthermore depends sensitively on additional effects not included in our simulation, such as the deformation of the DM halo~\citep{Kahlhoefer:2013dca,Kim:2016ujt}.

Nevertheless, to good approximation the position of the sub-cluster is independent of how exactly we model the galaxies in our simulation. We can therefore remove the uncertainties related to the centroid position of the sub-cluster by studying the difference between the separation observed in the two simulations:
\begin{equation}
 \overline{\Delta x}_1 - \overline{\Delta x}_2 = \overline{x_\text{gal,1}} - \overline{x_\text{gal,2}} \equiv \overline{\Delta x_\text{gal}} \; ,
\end{equation}
where the indices 1 and 2 refer to the simulations without and with the effects of DM self-interactions, respectively. The quantity $\overline{\Delta x_\text{gal}}$ is ideally suited for studying the effect of DM self-interactions on the trajectories of galaxies in merging galaxy clusters.

\begin{figure}
\centering
\includegraphics[scale = 0.7]{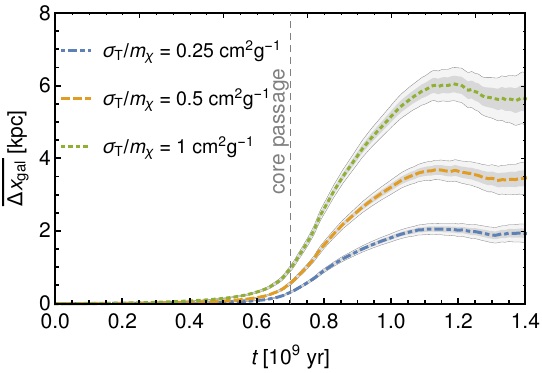}\\
\label{fig:deltaxgal_iso}
\vspace{2mm}
\includegraphics[scale = 0.75,clip,trim = 0 0 -18 0]{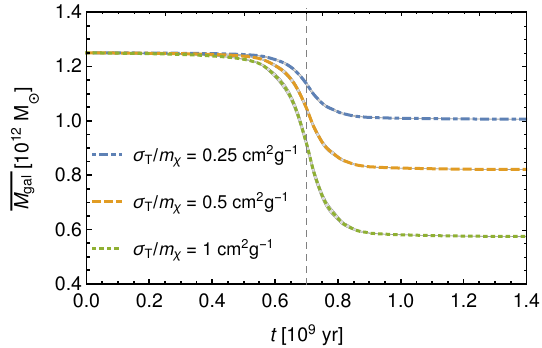}
\caption{Results from the numerical simulation of an isotropic DM self-scattering cross section for different values of $\sigma_{\text{T}}/m_{\chi}$. The top panel shows the time evolution of the difference between the average position of collisionless and collisional galaxies. The bottom panel shows the time evolution of the average mass of the collisional galaxies. 
The dark (light) grey bands indicate the statistical uncertainties from the finite sample size at the $1\sigma$ ($2\sigma$) level.}
\label{fig:result_iso}
\end{figure}

Let us first consider the isotropic case for different values of $\sigma_{\text{T}}/m_{\chi}$. We expect the largest effects to occur during and right after core passage, when both the velocity of the sub-cluster and the background density of the main cluster are largest. In our simulation this core passage occurs at $t_{\text{core}}\sim 0.7\cdot 10^9\text{yr}$. Indeed, Fig.~\ref{fig:result_iso} shows that the two quantities of interest, $M_\text{gal}$ and $\Delta x_\text{gal}$, remain essentially constant up to this point and then exhibit significant changes due to the presence of self-interactions. The grey bands in the top panel represent the uncertainties in the averaging resulting from the finite number of simulated galaxies~-- in the bottom panel this uncertainty is smaller than the thickness of the line.

For $\sigma_{\text{T}}/m_{\chi}=1 \mathrm{cm^2 \, g^{-1}}$, on average more than half of the mass of each galaxy evaporates due to the scattering of DM particles. The tail of escaping particles (as well as scattered particles that do not escape) lead to a significant deceleration of the DM halo. Compared to the simulation with collisionless galaxies, the galaxy centroid position in the simulation with an effective description of DM self-interactions is shifted by more than $5\:\text{kpc}$. This shift is clearly not a negligible effect in comparison to the total offset expected in such cluster collisions ($10\text{--}40\:\text{kpc}$), demonstrating that it is not a good approximation to treat the individual galaxies as collisionless test masses. Unsurprisingly, we find that the inclusion of DM self-interactions tends to shift the position of the galaxy centroid towards the back of the sub-cluster (to smaller values of $x$), thus reducing the galaxy-halo separation. The conclusion is that previous simulations of major mergers in the literature have likely overestimated the effect of DM self-interactions on the galaxy-halo separation.

The difference $\Delta x_\text{gal}$ between the average galaxy position in the two simulations peaks around $0.4 \cdot 10^9 \:\text{yr}$ after the core passage. At later times, the galaxies that have remained bound to the sub-cluster return to their equilibrium positions and therefore $\Delta x_\text{gal}$ decreases. Similarly, the skewness of the galaxy distribution peaks shortly after the core passage and then decreases again. The rate of this decrease depends on the precise definition of the ensemble average of the galaxy distribution, with tighter cuts on $|x_{\text{sub}}-x_{\text{gal}}|$ leading to more rapid relaxation. Crucially, the increase of $\Delta x_\text{gal}$ immediately after the core passage is independent of the precise definition of the ensemble average.

Let us now turn to the case of anisotropic scattering, as introduced in eq.~(\ref{eq:longrange}). As discussed in Sec.~\ref{sec:rates}, in order to compare cross sections with different angular dependence (parametrized by the cut-off velocity $w$) in a meaningful way, we need to specify a matching condition. Here we make the same choices as for Fig.~\ref{fig:testparticle}, i.e.\ we consider $w = 400,\ 800,\ 1600\,\mathrm{km\,s^{-1}}$ and determine the parameter $\sigma_0$ from the requirement $\sigma_\mathrm{T}(v_\text{ref} = 2000\,\mathrm{km\,s^{-1}})/m_\chi = 1 \mathrm{cm^2 \, g^{-1}}$. Since during core passage $v_\text{gal} \approx v_\text{sub} > 3000\,\mathrm{km\,s^{-1}} \gg w$ for all chosen values of $w$, large-angle scattering will be significantly suppressed. As a result, we expect smaller evaporation rates for the anisotropic case than for the isotropic case. The probability for small-angle scattering, on the other hand, remains sizeable and therefore we anticipate DM self-interactions to significantly affect the galaxy trajectories also in the anisotropic case.

\begin{figure}
\centering
\includegraphics[scale =  0.7]{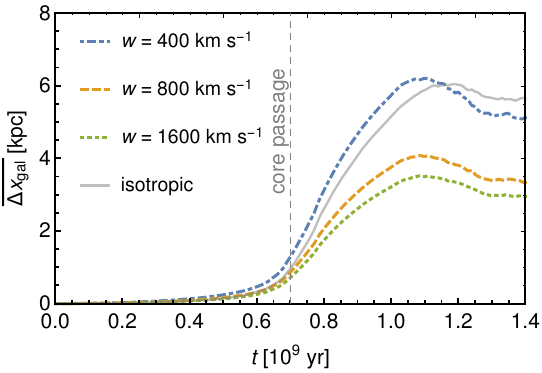}\\
\vspace{2mm}
\includegraphics[scale =  0.75,clip,trim = 0 0 -18 0]{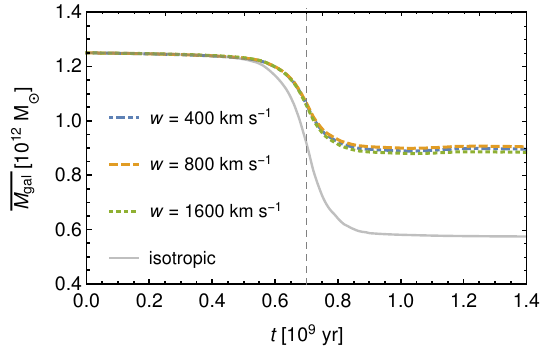}
\caption{Time evolution of the average spatial distance between collisionless and collisional galaxies (top) and time evolution of the average mass of the collisional galaxies (bottom) for anisotropic cross sections with different values of $w$.}
\label{fig:result_aniso}
\end{figure}

Our results are presented in Fig.~\ref{fig:result_aniso}. We observe that the mass fraction lost by the galaxies due to evaporation is very similar for the three choices of $w$ and amounts to about $30$ per cent over the course of the entire simulation, compared to a mass loss from evaporation of over $50$ per cent for the isotropic case. The average position of the galaxies, on the other hand, exhibits a clear dependence on $w$. Smaller values of $w$ lead to larger deceleration rates and therefore larger differences $\overline{\Delta x_\text{gal}}$ between the simulations without and with DM self-interactions. As for the case of isotropic scattering, $\overline{\Delta x_\text{gal}}$ grows after core passage and reaches maximum values of between $3 \, \mathrm{kpc}$ (for $w = 1600\,\mathrm{km \, s^{-1}}$) and $6 \, \mathrm{kpc}$ (for $400\,\mathrm{km \, s^{-1}}$).

The crucial observation is that the deceleration rate and the evaporation rate depend in different ways on the underlying particle physics. Figure~\ref{fig:result_aniso} shows three different particle physics models of SIDM, which lead to very similar evaporation rates but affect the motion of galaxy-sized DM haloes in very different ways. Conversely, the values of $\overline{\Delta x_\text{gal}}$ found for $w = 400\,\mathrm{km \, s^{-1}}$ are very similar to the ones in the isotropic case, while the evaporation rates are completely different in the two cases. We conclude that the momentum transfer cross section alone is insufficient to capture the effects of DM self-interactions and that the angular dependence is more accurately captured by the evaporation fraction $\chi_\mathrm{e}$ and the deceleration fraction $\chi_\mathrm{d}$ introduced above.

\section{Discussion}
\label{sec:discussion}

In this work we have derived an effective description of DM self-interactions for small DM haloes moving through a DM background. Our main results are the deceleration and evaporation rates in eqs.~(\ref{eq:decel}) and~(\ref{eq:evap}), with the deceleration and evaporation fractions $\chi_\mathrm{d}$ and $\chi_\mathrm{e}$ defined in eqs.~(\ref{eq:chid}) and~(\ref{eq:chie}). Crucially, we find that $\chi_\mathrm{d}$ and $\chi_\mathrm{e}$ capture different aspects of the underlying particle physics (roughly speaking the fraction of small-angle scatters and large-angle scatters, respectively), which are not captured by the momentum transfer cross section $\sigma_T$ alone. For example, if DM self-interactions have a strong preference for small-angle scattering, as in the case of long-range interactions, the deceleration rate can be significantly enhanced relative to the evaporation rate.

To illustrate this behaviour, we have considered a number of different particle physics parameters, which yield the same momentum transfer cross section (at a given reference velocity) but exhibit very different dependence on the scattering angle and the relative velocity. We show that evaporation can be significantly suppressed relative to the isotropic case, whereas the deceleration can be comparably large or even larger. The effective description is most accurate if the velocity of the DM halo relative to the DM background is large compared to the velocity dispersion of the background distribution, but even a sizeable velocity dispersion can be included in the effective description, leading to an enhanced evaporation rate and a reduced deceleration rate.

Our central focus has been on the behaviour of galaxies in the collision of two galaxy clusters. A particularly interesting question in this context is whether DM self-interactions can lead to a sizeable displacement of the galaxy distribution from its equilibrium position at the centre of the infalling DM halo. Clearly, in this context it is essential to model~-- at least approximately~-- the effect of DM self-interactions on the DM haloes of each individual galaxy. We quantify the magnitude of this effect using a simplified one-dimensional numerical simulation of an unequal merger like the Bullet Cluster. The effect of SIDM on the galaxies is found to be not negligible, implying that previous studies have likely overestimated the galaxy-halo separation in major mergers by simulating galaxies as collisionless test particles. The effective description of DM self-interactions that we propose is well-suited for an implementation in future numerical simulations of such systems in order to obtain more accurate results.

The expressions for $\chi_\mathrm{d}$ and $\chi_\mathrm{e}$ have been derived under the assumption that each DM particle scatters only once as it crosses the DM halo under consideration. This assumption is typically a good approximation for small DM haloes, but it may break down in the case of scattering with a strong angular dependence. In this case one enters the regime of \emph{frequent self-interactions}. As shown by \cite{Kahlhoefer:2013dca}, one then obtains evaporation and deceleration rates that are formally identical to eqs.~(\ref{eq:decel}) and~(\ref{eq:evap}), except that $\chi_\mathrm{d}$ and $\chi_\mathrm{e}$ now have to be calculated by averaging over a large number of scatters. Doing so, one finds that $\chi_\mathrm{d}$ and $\chi_\mathrm{e}$ are essentially independent of the astrophysical properties of the DM halo, leading to a universal drag force of the form $F_\text{drag} \propto \sigma \, \rho \, v^{2m}$, where $m = -1$ for long-range interactions and $m = 1$ for velocity-independent interactions. The effective description of DM self-interactions derived in this work can therefore straightforwardly be extended to frequent self-interactions, although the consistent implementation of such a drag force in numerical simulations remains challenging and is therefore left for future work.

In conclusion, we emphasize that galaxy-halo separations in merging galaxy clusters remain a highly promising signature of DM self-interactions. In particular, we have shown that the predicted effects depend in non-trivial ways on the underlying particle physics, so that an observation may enable us to infer information not only on the magnitude but also on the microscopic details of the DM self-scattering. Observables based on systems close to equilibrium, such as core formation in isolated DM haloes, on the other hand, may be more sensitive to the magnitude of the self-interactions but provide less information on the fundamental interactions of individual DM particles. Clearly, both avenues of studying SIDM should continue to be pursued in parallel. 

\section*{Acknowledgements}

We thank Torsten Bringmann, Marcus Brueggen, Richard Massey and Subir Sarkar for helpful discussions and Surhud More, Andrew Robertson and Andrew Benson for valuable comments on the manuscript. This work is supported by the DFG Emmy Noether Grant No. KA 4662/1-1 as well as the ERC Starting Grant `NewAve' (638528).




\appendix

\section{Realistic particle distributions}

\label{app:assumptions}

For the derivation of the evaporation and deceleration rates, eqs.~(\ref{eq:chie}) and~(\ref{eq:chid}), we have assumed that all incoming DM particles have the same direction and velocity and that all DM particles bound to the halo are in rest at the centre of the halo. In this appendix we review these assumptions and show that only small modifications of eqs.~(\ref{eq:chie}) and~(\ref{eq:chid}) are necessary to capture more realistic situations. These modifications are included in our numerical results in Section~\ref{sec:simulations}.

Let us first review the assumption that all incoming DM particles have the same velocity $v_0$. In practice we expect the incoming particles to have a velocity dispersion $v_\text{disp}$ in addition to the bulk velocity $v_0$, i.e.\ we define
\begin{align}
\mathbf{v}_0 & = \langle \mathbf{v} \rangle \\
v_\text{disp} &= \langle (\mathbf{v}_0 - \mathbf{v})^2 \rangle^{1/2} \; ,
\end{align}
where the brackets denote averaging over the DM velocity distribution $f(\mathbf{v})$:
\begin{equation}
 \langle x(\mathbf{v}) \rangle = \int f(\mathbf{v}) \, x(\mathbf{v}) \mathrm{d}^3 v \; .
\end{equation}

\begin{figure}
\centering
\includegraphics[scale = 0.51]{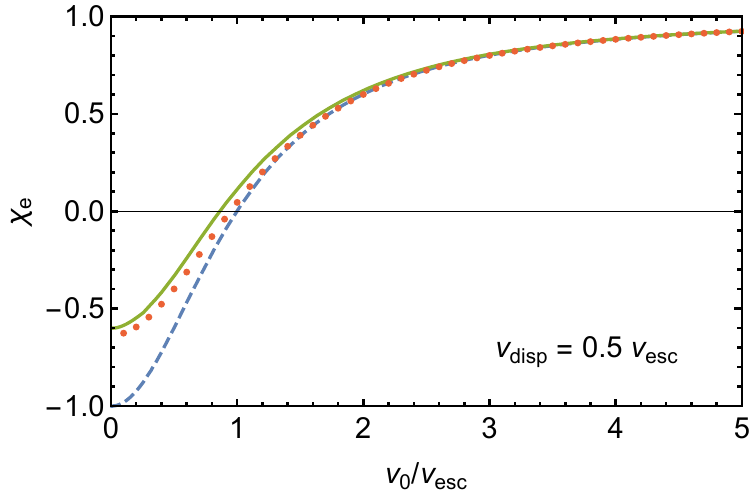}\\
\vspace{2mm}
\includegraphics[scale = 0.49,clip,trim=-10 0 0 0]{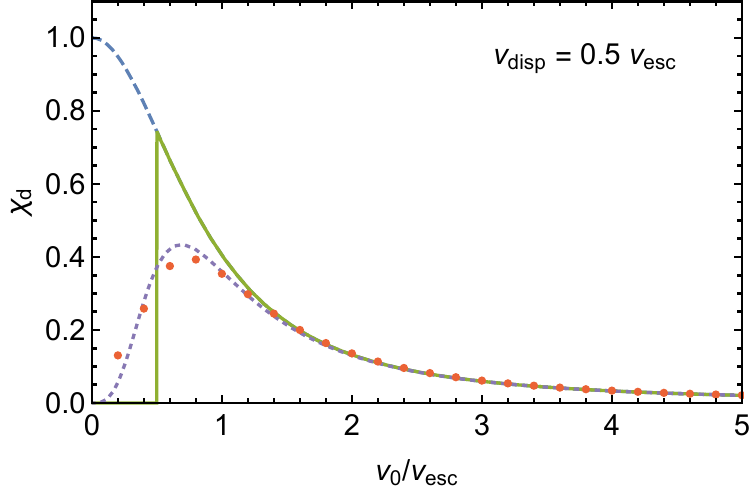}\\
\vspace{2mm}
\includegraphics[scale = 0.53,clip,trim=-13 0 0 0]{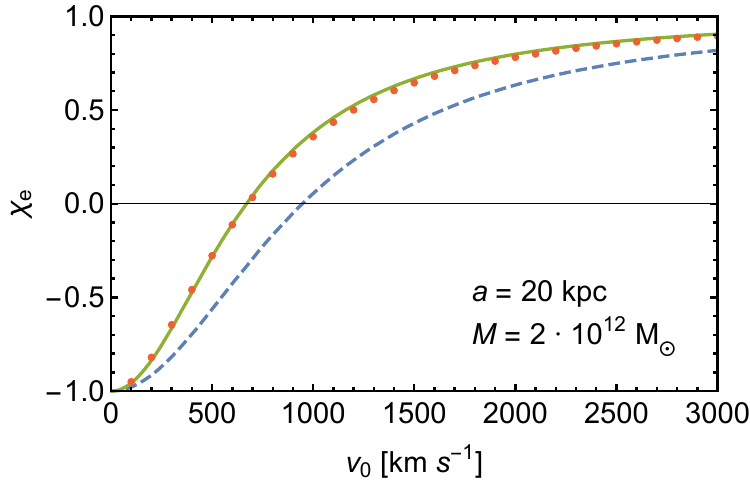}
\caption{Comparison of numerical estimates and simplified descriptions of the effect of realistic particle distributions. Top: evaporation rate including a Gaussian velocity distribution for incoming particles. Middle: deceleration rate including a variation in direction of the incoming particles. Bottom: evaporation rate accounting for the variation of the escape velocity. In each panel, the red dots indicate the numerical estimates, the blue dashed lines indicate the model prediction when not accounting for the respective effects and the green solid lines correspond to the proposed simplified description. In the middle panel, an alternative simplified description is shown in purple (dotted). See text for details.\label{fig:averaging}}
\end{figure}

If we assume for example a Gaussian velocity distribution, $f(v) \propto \exp \left[ -3 (\mathbf{v} - \mathbf{v}_0)^2 / (2 v_\text{disp}^2) \right]$, we find that the average velocity of incoming DM particles increases from $v_0$ to $\langle |\mathbf{v}| \rangle = (v_0^2 + v_\text{disp}^2)^{1/2}$ (see also app.~A of~\cite{Kim:2016ujt}). As a result, we expect a slight increase in the averaged evaporation rate $\langle \chi_e \rangle$. This change can be approximately captured by
\begin{align}
\langle \chi_e(|\mathbf{v}|) \rangle & \approx \chi_e( \langle |\mathbf{v}| \rangle) = \chi_e((v_0^2 + v_\text{disp}^2)^{1/2}) \; .
\end{align}
We confirm that this is indeed a good approximation by calculating $\langle \chi_e\rangle$ numerically for the case of isotropic scattering and setting $v_\text{disp} = v_\text{esc} / 2$ (see top panel of Fig.~\ref{fig:averaging}). While $\chi_e(v_0)$ (blue dashed line) underestimates the evaporation rate for small $v_0$, $\chi_e((v_0^2 + v_\text{disp}^2)^{1/2})$ (green line) gives a good description of the numerical data. We conclude that the velocity dispersion of the incoming DM particles can approximately be taken into account by adding it to their bulk velocity in quadrature.

For the deceleration rate the situation becomes more complicated, because not only the magnitude of the velocity of the incoming DM particles change, but also their direction. In the extreme case, where the velocity dispersion is large compared to the bulk velocity, $v_\text{disp} \gg v_0$, we would expect that the momentum transferred to the DM halo averages to zero. To confirm this expectation, we consider a simplified distribution of the form $f(\mathbf{v}) \propto \delta(|\mathbf{v} - \mathbf{v_0}| - v_\text{disp})$, i.e.\ we assume that all DM particles have a proper velocity of $v_\text{disp}$ relative to the bulk velocity $\mathbf{v}_0$ with isotropically distributed direction. For this distribution, we can then numerically calculate $\langle \chi_d(\mathbf{v}) \rangle$.\footnote{Due to the assumed symmetry of the velocity distribution, the deceleration must point in the same direction as $\mathbf{v}_0$, so we can still treat it as a scalar quantity.} 

As an example we show in the middle panel of Fig.~\ref{fig:averaging} the deceleration rate as a function of $v_0 / v_\text{esc}$ for $v_\text{disp} = v_\text{esc} / 2$. As expected, we find that the deceleration rate deviates significantly from the case with $v_\text{disp} = 0$ (blue, dashed) as soon as $v_0 \lesssim v_\text{disp}$. While it is difficult to model this turnover in detail, we can approximately capture the effect by setting $\chi_d(v_0) = 0$ for $v_0 < v_\text{disp}$ and neglecting the effect of the velocity dispersion for larger velocities (green). For a more accurate description of the numerical results one can multiply $\chi_d$ with a smooth function that changes from 0 to 1 around $v = v_\text{disp}$. For example, we find that replacing $\chi_d(v)$ by $\chi_d(v) \cdot v^3 / (v^3 + v_\text{disp}^3)$ yields a good fit to the numerical results (purple, dotted). Crucially, we conclude that no significant modification to $\chi_d(v_0)$ is necessary for $v_0 \gg v_\text{disp}$.

In addition to the velocity dispersion of the incoming DM particles, a realistic treatment should also include the velocities of the DM particles in the halo (which we have set to zero in the derivation of $\chi_d$ and $\chi_e$ above). To a first approximation, this effect can be included by defining the effective velocity dispersion $v_\text{disp,eff} = (v_\text{disp,1}^2 + v_\text{disp,2}^2)^{1/2}$, where $v_\text{disp,1}$ and $v_\text{disp,2}$ denote the velocity dispersion of the incoming DM particles and of the DM halo, respectively. However, for the situation that we are interested in, i.e.\ a small halo moving through the background density of a much bigger halo, one typically expects $v_\text{disp,1} \gg v_\text{disp,2}$ and it is therefore a good approximation to neglect $v_\text{disp,2}$.

A final assumption that we need to review is that all scatters occur in the centre of the DM halo. This assumption enters via the escape velocity, which is largest in the central region. Away from the centre, DM particles will be less tightly bound, so we expect evaporation to occur more readily. For a given DM density profile $\rho(r)$, we can calculate the escape velocity as a function of the radius $r$ via $v_\text{esc}(r) = \sqrt{- 2 \Phi(r)}$, where $\Phi(r)$ is the gravitational potential. We can then calculate the average escape velocity 
\begin{equation}
\bar{v}_\text{esc} = \frac{1}{M} \int \rho(r) v_\text{esc}(r) \mathrm{d}^3 r  \; ,
\end{equation}
where $M$ denotes the total mass of the DM halo. For example, for a Hernquist profile with scale radius $a$, we find that $\bar{v}_\text{esc} \approx 1.03 \sqrt{G \, M / a} \approx v_\text{esc}(a)$. We then expect that the density-averaged evaporation and deceleration rates $\bar{\chi}_{e,d}$ can be approximated by calculating the respective rate for the average escape velocity: $\bar{\chi}_{e,d} = \chi_{e,d}(\bar{v}_\text{esc})$. To confirm this expectation, we explicitly calculate $\bar{\chi}_e$ for a Hernquist profile with $M = 2 \cdot 10^{12} \text{M}_{\odot}$ and $a = 20\:\text{kpc}$ as a function of $v_0$. As shown in the bottom panel of Fig.~\ref{fig:averaging} the results are well described by $\chi_{e}(\bar{v}_\text{esc})$, while taking $\chi_{e}(v_\text{esc}(r = 0))$ would significantly underestimate the evaporation rate.

To conclude this discussion we note that an additional complication may occur when considering arbitrary positions $\mathbf{x}$ of scattering events: If $\mathbf{x}$ and $\mathbf{v}$ are not parallel, the two scattering DM particles will no longer move on radial orbits. For a complete treatment one would then need to take into account the full three-dimensional distribution of scattering angles and calculate the trajectories of the outgoing DM particles, which would clearly require a Monte Carlo simulation. Nevertheless, as long as $\rho(r)$ and $f(|\mathbf{v} - \mathbf{v}_0|)$ are isotropic, the symmetry of the problem ensures that the only net effect can be in the direction of $\mathbf{v}_0$, so that our simplified treatment is likely a good approximation.

\section{Galaxy distribution functions}
\label{app:1deddington}

A central challenge of running a one-dimensional toy simulation of galaxies moving inside a DM halo is to find a suitable distribution function $f(x, v)$ that can be used to randomly generate the initial positions and velocities for the desired number of galaxies. The fundamental requirement for such a distribution function is that it should be time-independent, i.e.\ that the distribution of galaxies at any later point in time should still be given by $f(x, v)$ (as long as the DM halo does not experience acceleration). In practice that means that $f(x, v)$ should only depend on quantities that are integrals of motion. In particular, $f(x ,v)$ is certainly time-independent, if it can be written as a function of the reduced energy $\mathcal{E} \equiv \Psi - v^2/2$, where $\Psi(x) = - \Phi(x)$ with a given gravitational potential $\Phi(x) < 0$.

For given $f(x, v)$, we can calculate the density profile 
\begin{equation}
\rho(x) = \int_0^{v_\text{esc}} f(\mathcal{E}(x, v)) \mathrm{d}v \; , 
\end{equation}
where the escape velocity depends on $x$ via $v_\text{esc} = \sqrt{2 \Psi(x)}$. It is important to note that, as long as we are interested in describing a component that only gives a sub-dominant contribution to the total mass of the system, there is no direct link between $\rho(x)$ and $\Psi(x)$, i.e.\ it is perfectly possible for the galaxies to have a density profile different from the one of DM. Making use of $\mathrm{d}\mathcal{E}/\mathrm{d}v = -v = -\sqrt{2 (\Psi - \mathcal{E})}$, we obtain
\begin{equation}
\rho(x) = \int_0^\Psi(x) \frac{f(\mathcal{E})}{\sqrt{2(\Psi - \mathcal{E})}} \mathrm{d}\mathcal{E} \; ,
\end{equation}
which makes explicit the fact that $\rho$ can be thought of as a function of $\Psi$. In analogy to the derivation of the Eddington formula in three dimensions~\citep{Eddington}, we can now solve the Abel integral equation for $f(\mathcal{E})$:
\begin{equation}
f(\mathcal{E}) = \frac{\sqrt{2}}{\pi} \int_0^\mathcal{E} \frac{\mathrm{d}\Psi}{\sqrt{\mathcal{E} - \Psi}} \frac{\mathrm{d}\rho(\Psi)}{\mathrm{d}\Psi} \; ,
\label{eq:1deddington}
\end{equation}
where we have made use of the fact that $\rho(\Psi = 0) = \rho(r \to \infty) = 0$.

Eq.~(\ref{eq:1deddington}) can be used to calculate $f(\mathcal{E})$ for given $\Phi$ and $\rho$. It is worth emphasizing that it is not guaranteed that the function obtained in this way is physical, i.e.\ that $f(\mathcal{E}) \geq 0$ everywhere. For the purposes of our simulation, we consider a very simple case, namely that $\Phi$ is given by the potential of a Hernquist profile (consistent with what we assume for the DM distribution in the halo) with total mass $M$ and scale radius $a$ and that the one-dimensional density profile is given by
\begin{equation}
 \rho(x) = \frac{m \, a}{(|x| + a)^2} \; ,
\end{equation}
where $m$ is the total amount of mass in galaxies. This density profile is finite for $r \to 0$, as necessary for a one-dimensional simulation, and asymptotes to the scaling $r^2 \rho \propto 1/r^2$ of the three-dimensional Hernquist profile at large distances. From eq.~(\ref{eq:1deddington}) we then obtain
\begin{equation}
 f(\mathcal{E}) = \frac{8 \sqrt{2} \, a \, m \, \mathcal{E}^{3/2}}{3\pi \, G_\mathrm{N}^2 \, M^2} \; .
\end{equation}
This distribution function can be sampled numerically to generate the initial conditions for our simulations. We have confirmed explicitly that this distribution function is indeed time-independent for an isolated halo.


\bsp	
\label{lastpage}
\end{document}